\documentclass[doublecol]{epl2}
\title{Pattern formations driven by cyclic interactions: a brief review of recent developments}
\shorttitle{Pattern formations driven by cyclic interactions}

\author{A. Szolnoki\inst{1}, B.F. de Oliveira\inst{2}, and D. Bazeia\inst{3}}
\shortauthor{Szolnoki, de Oliveira, Bazeia}
\institute{\inst{1}Institute of Technical Physics and Materials Science, Centre for Energy Research, Hungarian Academy of Sciences, P.O. Box 49, H-1525 Budapest, Hungary\\
\inst{2}Departamento de F\'\i sica, Universidade Estadual de Maring\'a, 87020-900 Maring\'a, PR, Brazil\\
\inst{3}Departamento de F\'\i sica, Universidade Federal da Para\'\i ba, 58051-970 Jo\~ao Pessoa, PB, Brazi\\}

\abstract{Lotka's seminal work (\Name{Lotka A.~J.} \REVIEW{Proc. Natl. Acad. Sci. U.S.A.}{\bf 6}{1920}  410) ``on certain rhythmic relations'' is already one hundred years old, but the research activity about pattern formations due to cyclical dominance is more vibrant than ever. It is because non-transitive interactions have paramount role on maintaining biodiversity and adequate human intervention into ecological systems requires deeper understanding of related dynamical processes. In this perspective article we overview different aspects of biodiversity, with focus on how it can be maintained based on mathematical modeling of last years. We also briefly discuss the potential links to evolutionary game models of social systems, and finally, give an overview about potential prospects for future research.}

\begin{document}

\maketitle

\section{Introduction}

To reveal predator-prey-like relations between competing species is an essential task to understand the stability of ecosystems in our world \cite{ives_s07}. From this point of view non-transitive relations, where there is a certain cyclic dominance between competitors, offers not just an intellectual challenge, but also helps us to understand the stability of diversity in highly complex systems. Besides the well-known examples which are already surveyed in Ref.~\cite{szolnoki_jrsif14}, we only mention here some recently published works to enrich the list of those biological and ecological systems where the mentioned cyclic interaction is identified. Spiral waves in cell colonies \cite{grace_pcbi15}, oscillation in salmon \cite{schmitt_jtb14} or larvae populations \cite{rafikov_ec14}, cyclical succession in grazed ecosystems \cite{ruifrok_tpb15}, or social amoebas \cite{shibasaki_prsb18} are new exciting fields, but the list can be extended easily \cite{frey_18,kosakovski_pone18,das_jml19,liao_s19,lewin-epstein_prsb20}. The principal goal of our present paper is not just to summarize the recent developments of cyclically dominant systems obtained in the last years, but we also want to emphasize that non-transitive relations are not restricted to biological systems only. There are several situations in conflicts of human societies when the applied strategies of competitors can be modeled by a similar approach of evolutionary game theory.

We must note that several excellent reviews have been published in the last two decades which already discussed not just the basic problem in detail, but also surveyed the applied theoretical approaches including mean-field, reaction-diffusion, Langevin, Fokker-Planck, and Ginzburg-Landau equations \cite{szabo_pr07,frey_pa10,mowlaei_jpa14,wang_pr16,dobramysl_jpa18,cianci_pa14,cazaubiel_jtb17}. Hence for further technical details we suggest our readers to check these review papers, because in this perspective article our principal goal is to present a brief report about recent developments obtained in this rapidly developing field in the last couple of years. Our other goal is to provide some prospects about possible forthcoming research efforts.

Since our survey basically focus on pattern formations in spatially extended systems, we first summarize the two basic model approaches that are intensively applied for this goal. Afterwards we survey the recent key observations about the conditions which determine when biodiversity can be preserved and which are the components that jeopardize the coexistence of competing partners. Next we discuss the various extension of the basic model, which resulted in qualitatively new situations and phenomena that cannot be observed in the basic case. As mentioned, while the basic models were motivated by ``organic systems'', conceptually similar self-organizations and patterns can be observed in systems far from biology. We therefore summarize those related phenomenon that can be detected in other evolutionary game models. Last we close our brief survey with an outline of possible directions for future research to reach a deeper understanding of socio-ecological systems.

\section{Lotka-Volterra versus May-Leonard models}

In the simple spatial version of Lotka-Volterra ($LV$) model three species fight for space where every competitor beats another one in a cyclic way similarly to the well-known rock-scissors-paper game \cite{lotka_pnas20,volterra_31}. An important detail of the microscopic dynamics is that the dominant species occupies the position of the neighboring partner immediately, hence the total sum of individuals is conserved. Mobility of species can also be introduced here which is implemented through the swapping of individuals occupying neighboring sites of the interaction graph \cite{avelino_pre12b,nagatani_em17,hashimoto_jpsp18}.

In the alternative interpretation of cyclic dynamics, which was suggested by May and Leonard ($ML$), the competition is divided into two elementary steps \cite{may_siam75}. First, a selection may take place in a way that the invaded species goes extinct by leaving an empty room behind. This free position can be filled by a neighboring species via a reproduction step. Since both steps are stochastic in general, the total sum of individuals is not conserved anymore \cite{bazeia_epl17,pereira_ijmssc18,avelino_pre19,avelino_epl19}.

The most fundamental problem is whether the mentioned difference between microscopic rules results in a qualitative change in the system behavior or the actual effect is robust and remains valid independently of the details of microscopic dynamics. There is no unambiguous answer to this question because to extend the models to spatial systems cannot solve the discrepancy we already observed in well-mixed (or mean-field) conditions. In the latter case the solution of $LV$-equations is a neutral orbit, but $ML$-model predicts unstable spiral toward one of the trivial fixed points in the ternary diagram. On the other hand, in spatial systems there are examples where both approaches agree while there are also cases when they disagree about the qualitative feature of system behavior.

For example, both models agree that varying reaction rates have little effect on the dynamical evolution, or some fixed spatial disorder has only minor effect on species coexistence. Further notable progress was to explore the robustness of the so-called ``survival of the weakest'' effect. Accordingly, in case of heterogeneity the species which has the smallest invasion power benefits the most from the unequal invasion rates. This phenomenon remains valid both in $LV$ and $ML$ systems \cite{menezes_epl19,avelino_pre19b,depraetere_c18}, but has limited validity when the number of species is increased \cite{avelino_pre20}.

We must admit that the picture is even more complex, because an unequal invasion rate can be reached not only by changing the stochastic invasion rates between competitors, but also by introducing invasion in the reversed direction. While such modification has no particular consequence on a $LV$ model, the symmetry breaking of the invasion flow is proved to be a critical point in $ML$ systems, because it jeopardizes the coexistence of the three competing species \cite{bazeia_csf20}.

Another issue of current interest concerns the search for distinct descriptions of possible chaotic behavior of cyclically dominant systems. A new tool, called basin entropy concept \cite{daza_srep16} was suggested recently by Mugnaine~{\it et~al.} \cite{mugnaine_epl19}. This is different from the Hamming distance density \cite{hamming_bstj50} used before in Ref.~\cite{bazeia_epl17}, and may foster new lines of study.
 
\section{Fundamental conditions to maintain biodiversity}

The decisive role of interaction graphs on the stability of coexistence has already been known for almost 15 years \cite{szabo_jpa04}, and this effect was confirmed by several recent studies since then \cite{laird_o14,sato_k_mbe19}. Beyond model studies, real-life systems, like microbial communities illustrate that the stability of an ecological system depends strongly on the structural environment in which they reside \cite{lowery_pnas19}. An important factor is whether competitors have a chance of long range interaction, because the latter may destroy the stable coexistence if it exceeds a threshold value \cite{rulquin_pre14}.

Another crucial factor is the weight of mobility in microscopic dynamics \cite{reichenbach_n07,mobilia_g16}. Similarly to the long-range interactions, if it exceeds a certain value, biodiversity is jeopardized and lost. As we will refer in the rest of this paper, the mentioned mobility factor has been tested in several other models. Here we just note that directional mobility, when an individual moves in the direction with a larger number of possible selection targets, may also reduce the chance of diversity \cite{avelino_pre18}.

While the concept of zealots, agent with fixed states, was originally introduced for voter model \cite{mobilia_jsm07}, similar situations can be easily raised in cyclically dominated systems \cite{verma_pre15}. It turned out that they are effective in taming the amplitude of oscillations that emerge due to mobility and/or interaction randomness \cite{szolnoki_pre16}. Notably, other forms of heterogeneity, including spatial heterogeneity \cite{baker_jtb20}, site-specific invasion rates \cite{szolnoki_srep16b}, or turbulence \cite{groselj_pre15} have also been studied. As a general result, breaking the uniform environment assumption in the form of habitats that locally provide one of the species with an advantage, destabilizes spiral waves hence diversity. 

\begin{figure*}
\begin{center}
\includegraphics[width=17.8cm]{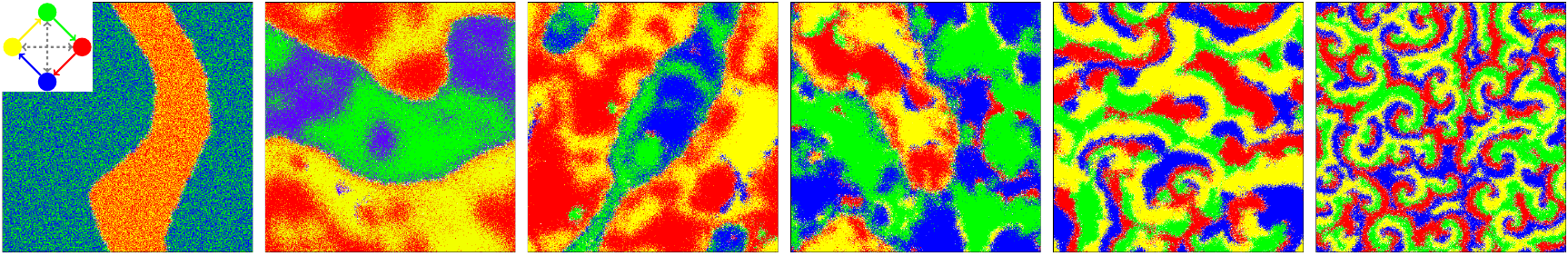}
\caption{\label{phases} Characteristic patterns of phases in dependence of inner bidirectional invasion rates between peer species \cite{bazeia_pre19}. The food-web of four competing species is shown in the inset of left panel. Here, beside the traditional cyclic loop, there is a bidirectional invasion between peer species, marked by dashed arrows. From left to right the strength of this latter invasion is changed only, resulting in significantly different solutions. This case illustrates nicely that the pure analysis based on the topology of food-web cannot be enough to predict the final solution reliably in cyclically dominated systems.}
\end{center}
\end{figure*}

An interesting extension was made by Bassler~{\it et~al.} who developed the original model by providing agents a ``social temperament'' to be either introverted or extroverted, hence leading them to cut or add links. Such kind of diversity stabilizes coexistence \cite{bassler_pre19}. Quantifying diversity has a paramount importance in general to handle chaotic systems evolving under the guidance of specific cyclic dynamics \cite{bazeia_srep17}. Here it is possible to relate the correlation length and the average density of maxima of the abundance of a typical species by using the Hamming distance concept \cite{hamming_bstj50} and the counting of conductance peaks in open quantum dots \cite{ramos_prl11}. Moreover, the comprehensive description and characterization of spiraling patterns gives interesting explanation for the lack of observation of spiraling patterns in some microbial experiments \cite{szczesny_pre14}.

We also note that Park and his collaborators have studied the possible role of nonlinear intraspecific competition in a series of papers \cite{park_sr17,park_c18e,park_c18d,park_amc18}. They found that nonuniform intraspecific competitions may act on biodiversity to surprisingly enlarge the spectrum of coexistence states that emerge and persist.

Needless to say, such a brief review cannot be complete and is unable to give account for all possible directions, like the consequences of mutant species \cite{mobilia_jtb10,park_c18}, environmental randomness \cite{west_pre18,west_jtb20}, the role of myopic strategy update \cite{varga_pre14}, or the case of metapopulation models \cite{nagatani_srep18,nagatani_jtb19,nagatani_pa19b,nagatani_jpsj20}. Hence, we refer the interested reader to the original works for further details.

\section{Extensions beyond three species}

While the basic idea of cyclic dominance was introduced in a three-member rock-scissors-paper game, it can be extended easily by more competitors. Classification of more general models can be made based on actual relation of competitors. Accordingly, in the so-called $(N,r)$ model we have $N$ competing species where everyone has $r$ preys in the food-web \cite{roman_pre13}. In Ref.~\cite{roman_jtb16} Pleimling and his collaborators suggested a matrix approach formalism where $N$ competing species define an $N \times N$ adjacency matrix. Here the matrix element $a_{ij}$ is equal to one if $i$ preys on $j$, and zero otherwise. This technique allows to predict partnership formation in a class of food-webs, which are based on binary interactions. This scenario is valid in several cases, but not always because there are situations which can only be understood by assuming multi-point interactions \cite{szolnoki_pre14c,hernandez_epjb14,kelsic_n15,szolnoki_njp15}. For instance, the presence of a third party, as a protection spillover, may influence the intensity of the original invasion within the traditional predator-prey pair. 

To be more specific, already the extension of the cycle to four members brings a qualitatively new concept of neutral partners. These members are not in a direct predator-prey relation \cite{intoy_jsm13}, which makes possible the formation of domains composed by the mentioned neutral partners \cite{intoy_pre15,bayliss_pd20}. Naturally, direct invasion can also be established between them where the invasion activity is not necessarily equal to the invasion rate in the main loop. In a model where such a heterogeneous food-web is defined, there is a transition from a coexistence state to one with extinction in dependence of the additionally introduced invasion rate \cite{lutz_jtb13}. Notably, bidirectional invasion among these peer, or otherwise neutral species provides an amazing series of different states by varying solely the intensity of this bidirectional invasion strength \cite{bazeia_pre19}. This wealth of phases is illustrated in Fig.~\ref{phases}.

The next step of extension is the five-species model, also called as rock-paper-scissors-lizard-spock game. Here all pair invasions are established between competitors, hence everyone has two predator and two prey partners. In this model a single $q$ control parameter is introduced as the ratio of the principal and next-nearest-neighbor invasion rates. In the vicinity of the golden ratio ($q=(\sqrt 5-1)/2$) the fluctuation becomes very large and the system evolves into a traditional three-species solution if the linear size is not large enough \cite{kang_pa13}. This finite-size effect is related to a phenomenon where the direction of invasions between emerging associations is reversed. The reported phenomenon illustrates nicely that the topology of the food web does not always determine the final state of the evolution \cite{vukov_pre13}. Similarly to the classic rock-scissors-paper model, the mobility here also plays a decisive role. As it is increased, the five-species state changes into a three-species coexistence state and by further increasing the strength of mobility the system evolves onto a uniform, single-species state \cite{cheng_hy_srep14}. However, as Park and Jang pointed out, strong asymmetry of spontaneous (cyclic) and alternative (next-nearest neighbor) competitions can lead to robust coexistence even if mobility is high \cite{park_c19b}.

Evidently, increasing the number of species further, by studying six \cite{esmaeili_pre18} or even nine species \cite{brown_pre19} makes the basic model more challenging, where novel approaches, like monitoring the system response after a sudden external perturbation, could be useful to get a deeper insight into the nonequilibrium dynamics.

To close this subsection we briefly mention an independent way of model extension when the so-called apex predator is introduced into the traditional cyclically dominant loop \cite{wallach_o15}. From ecological viewpoint the presence of such predator influences the stability of an ecosystem greatly and the related model study revealed that it spreads uniformly in space, and contribute to destroy the traditional spiral patterns \cite{postlethwaite_epl17}. In this way they keep biodiversity by diminishing the average cluster size of cyclically competing species \cite{souza-filho_pre17}.
Furthermore, in the presence of such predator the classical species experience higher growth rates, and lower competitive death rates than they do compared to the usual three-species case \cite{bazeia_epl18}.

\subsection{Competition of alliances}

As we mentioned, the presence of more than three species may change the evolutionary dynamics significantly because neutral species may emerge who have no direct interaction in the absence of predator-prey relation. Consequently, in an extended model the basic problem is not necessarily reduced to the simple question as whether competitors coexist or the evolution terminates into a homogeneous state. Instead, a new kind of solutions may emerge which are formed by a subset of available species. Such a smaller group of competitors can behave like an alliance and keep external invaders away from their territory. Interestingly, the mentioned alliance may be formed by direct competitors who are in a proper predator-prey relation to each other \cite{cheng_hy_srep14}, or in certain cases the above mentioned neutral pairs may also compose such a unit \cite{dobrinevski_pre14,avelino_pre14}. 

\begin{figure}
\begin{center}
\includegraphics[width=8.0cm]{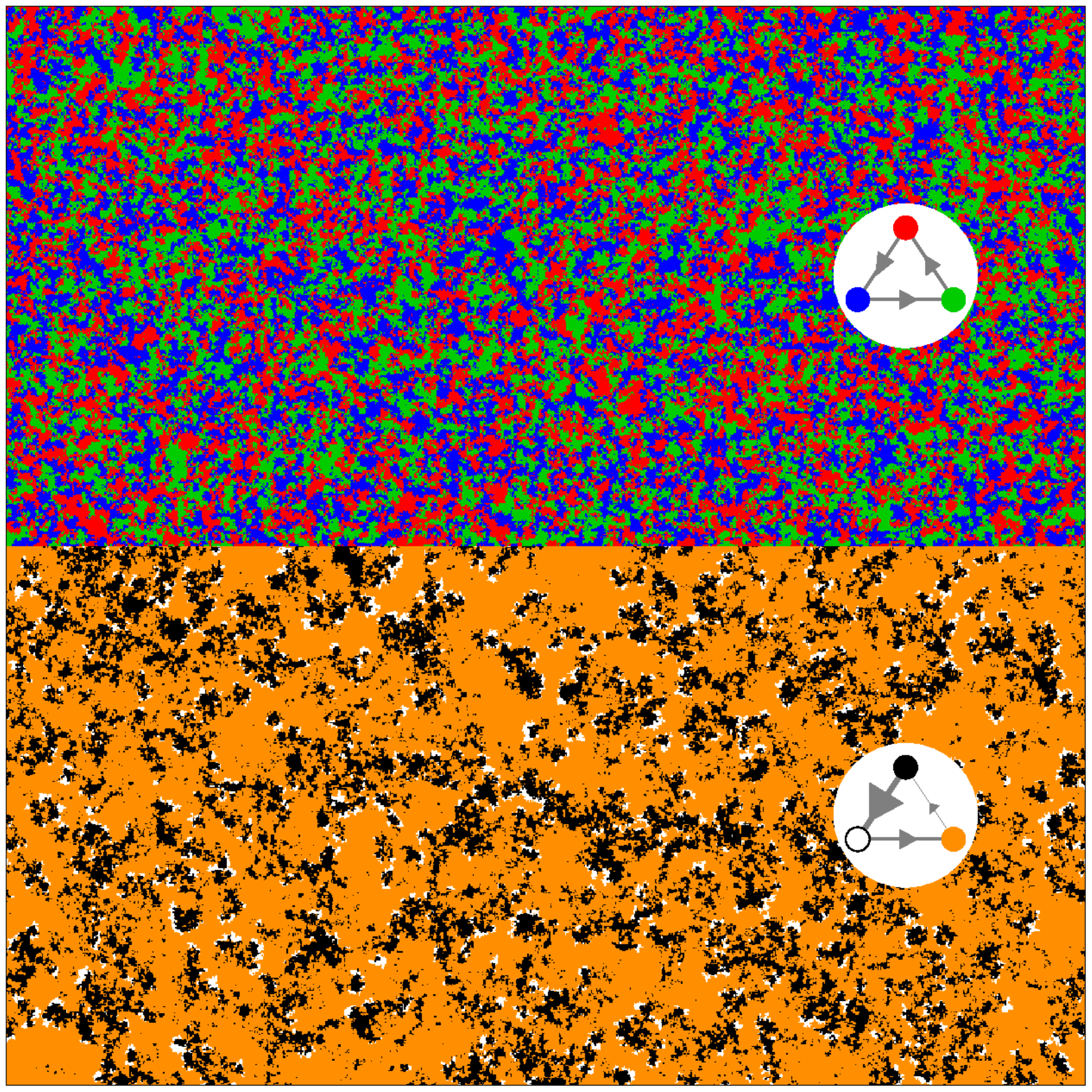}
\caption{\label{prep} Prepared initial state in a six-specie $LV$ model where two sub-solutions can emerge without disturbing each other. Competing species are marked by different colors, as shown by the insets. Notably, both in the top and the bottom half of available space cyclically dominant species form a stable solution. In the bottom half, however, the invasion rates between partners are heterogeneous, as indicated by the sizes of invasion arrows. The unequal invasion rates result in a highly heterogeneous pattern in the bottom half. When competition is released between the solutions then the latter becomes vulnerable against the invasion of the other solution where equally strong partners form the alliance \cite{blahota_epl20}.}
\end{center}
\end{figure}

The mentioned description practically means that not species, but solutions compete for space. Consequently, the traditional and frequently applied simulation technique, which is launched from a randomly distributed state, may result in a misleading conclusion because it offers non-equal chance for all possible solutions. It means that a certain solution needs more time and space to emerge and this chance has not necessarily granted when the simulation is started from a random initial state especially at a small system size. This may cause a serious finite-size effect where the final state depends sensitively on the initial state of simulation \cite{kim_c17}. Naturally, if the system size is large enough then there is a chance for every solution to emerge somewhere in the space and the more vital one can eventually invades the whole system. But this system size may not always be feasible for a simple simulation, hence it does not always find the proper solution. To overcome this difficulty we may need to apply the so-called prepared initial states where different solutions may emerge separately in space by not allowing invasion through the border lines. Afterwards, when they reach their stationary states, we open the borders and monitor which solution dominates the other one. An example is shown in Fig.~\ref{prep}. Here a six-member $LV$ model is studied, where besides the global cycle, members with odd or even labels also form smaller cycles of dominance. Importantly, in one of the sub-cycles we apply heterogeneous inner invasion rates. Interestingly, this has a serious negative impact on the vitality of the mentioned alliance against the one where equally strong partners form the alternative sub-cycle \cite{blahota_epl20}.

In another intensively studied case the sub-solutions are equally strong and the evolution of interfaces separating them is driven by its curvature \cite{avelino_pre12b,pereira_ijmssc18,avelino_pre19,mitarai_pre16}, similarly to the ordering kinetics of Ising model \cite{bray_ap94}. This parallel becomes more apparent in higher dimensions where the usage of string networks reveals scaling law coarsening dynamics analogous to that found in condensed matter physics and cosmology \cite{avelino_pla14,avelino_pla17}. There is, however, a significant difference. Rotating spirals may have influence on the coarsening process, yielding a time exponent different from the standard one \cite{brown_pre17,esmaeili_pre18}. Evidently, as we increase the number of competitors then more complex patterns, like spirals within spirals  can be observed \cite{brown_pre19}.

\section{Evolutionary game strategies forming cyclic dominance}

As we already stressed earlier, the evolutionary dynamic driven by cyclic dominance is more widespread and several other examples can be found beyond biological systems. The first identified and intensively studied case is the relation of cooperator, defector, and loner strategies in social dilemmas \cite{hauert_s02,jiang_pre18,canova_jsp18}. Here a loner refuses being exploited by defectors, but they collect lower income than cooperators do, which establishes the cyclic dominance. It is worth noting, however, that this effect may be sensitive to the use of microscopic dynamics and remains invisible if players make a fully rational decision during their strategy updates. This frustration, however, can be solved automatically in a diluted graph where the possible movement of agents diminishes the artifact consequences of some microscopic rule \cite{cardinot_njp19}.

Naturally, introducing more sophisticated strategies, such as rewarding cooperators \cite{wu_y_srep17,cheng_f_amc20}, punishing defectors \cite{szolnoki_pre11b,perc_sr15,liu_jz_csf18}, peer exclusion \cite{quan_j_c19}, or the presence of risk-averse hedgers \cite{guo_h_jrsif20} gives a chance to establish more subtle relations between competing strategies. In these complex systems the emergence of cyclic dominance seems to be more natural than an exception. 

The parallel with traditional cyclically dominated models is not just a general theoretical option, but yields a strong practical link. More precisely, what we learn from ``sterile" or abstract $LV$ and $ML$ systems may be utilized directly in models motivated by social problems. To illustrate this, our first example is the impact of inner invasion speed on the vitality of an alliance based on cyclic dominance. In an $LV$ system it was already pointed out that when two otherwise equally strong alliances compete then the one which applies faster inner invasion can beat the other alliance \cite{perc_pre07b}. In the classic social dilemma, when we introduce informed strategies who are willing to pay an extra cost for learning the strategy of opponents then two competing cyclic loops can be identified among the available strategies. Importantly, the effective strength of invasion within these loops can be tuned by changing the values of the payoff element. As a result, the dominance between the mentioned loops can be changed by solely varying the payoff values and a weaker strategy in a faster rotating triplet can thus overcome an individually stronger competitor \cite{szolnoki_epl15}.

Our other example is the already mentioned ``survival of the weakest effect'' which was proved to be robust both in $LV$ and $ML$ systems \cite{menezes_epl19,avelino_pre19b,depraetere_c18}. A significant difference between cyclically dominant models and social systems is that the invasion strength cannot be adjusted as directly as in a rock-scissors-paper type model. Instead, we may introduce a strategy-dependent learning activity, which result in a similar change in the invasion flow around strategies. Crucially, this intervention into the original model will not only lower the strategy adoption between the predator-prey strategies, but potentially it also modifies the strategy adoption from the reversed direction when a player of the actual strategy wants to imitate the strategy of a third party. Despite this conceptual difference, the basic feature of the survival of the weakest effect remains intact. More specifically, the lowering of the effective strategy adoption between a ``predator'' and ``prey'' strategies will promote the growth of former population in the presence of a cyclic invasion \cite{szolnoki_csf20b}.

\begin{figure}
\begin{center}
\includegraphics[width=4.1cm]{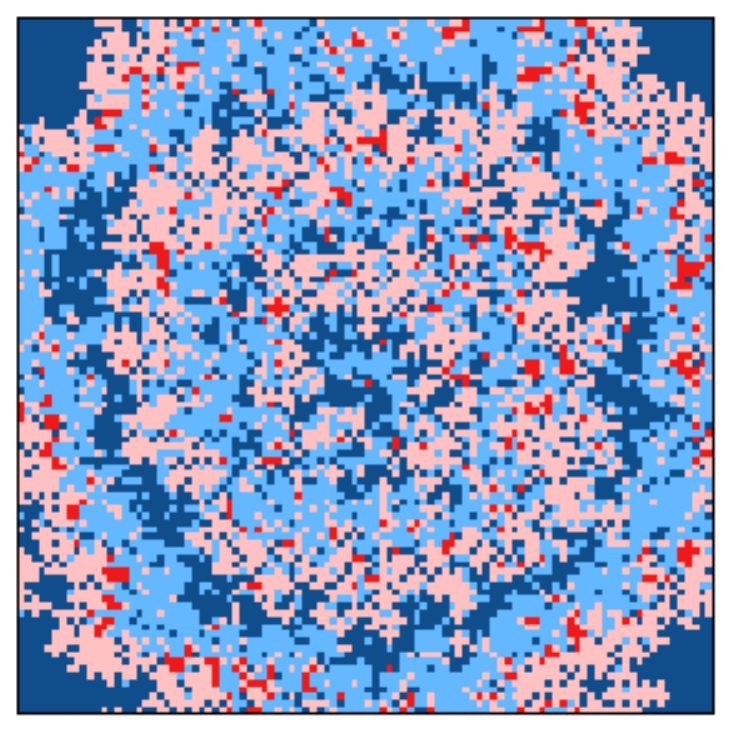}\includegraphics[width=3.92cm]{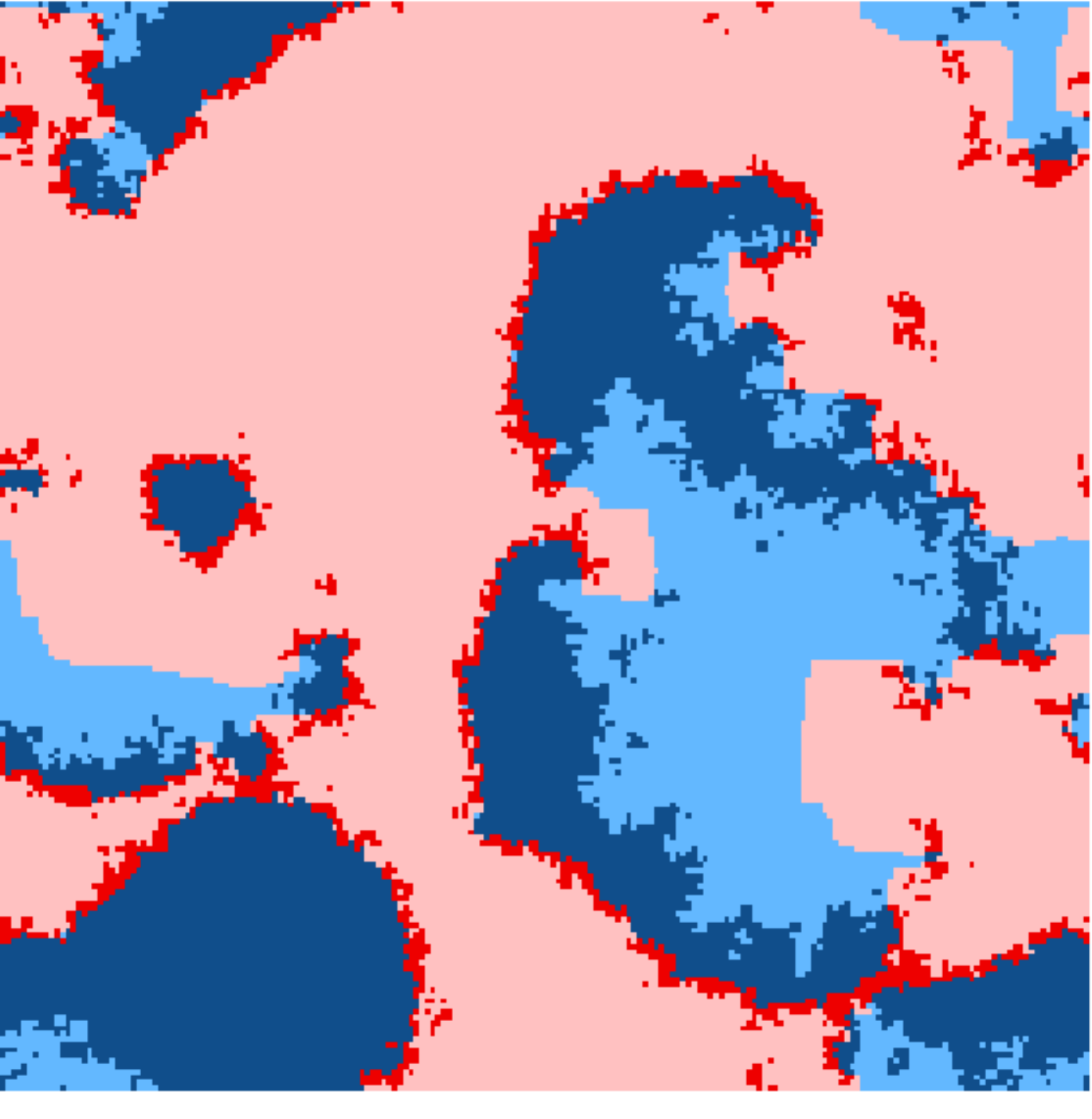}
\caption{\label{CD} Patterns driven by cyclic dominance in two-strategy models. In both cases different shades of blue and red colors denote different status of cooperator and defector strategies respectively. In the left panel light colors denote strategies with minimal, while dark colors mark players with maximal strategy adopting capacity \cite{szolnoki_pre10}. In the other case, shown in the right panel, both strategies apply imitation (light colors) or myopic best response (dark colors) strategy update rule \cite{danku_epl18}.}
\end{center}
\end{figure}

Finally we note that to observe a dominance we may not need to have three or more independent strategies. For example, if we have two strategies only, but the state of a given strategy, which determines the dynamical rule, depends on time then we can detect similar propagating fronts as previously reported for cyclic predator-prey models \cite{szolnoki_pre10}. A typical pattern of propagating fronts is shown in the left panel of Fig.~\ref{CD}. A conceptually similar cyclical competition was reported by L{\"u}tz~{\it et~al.} in Ref.~\cite{lutz_g17}. Alternatively, players may use different ways to update their strategies, which offers another degree of freedom. Consequently, multi-states may emerge no matter we only have two basic strategies. At certain parameter values they dominate each other cyclically and produce the well-known invasion fronts with rotating spirals, as it is illustrated in the right panel of Fig.~\ref{CD} \cite{danku_epl18}.

\section{Prospects}

While the literature has a huge amount of articles that study cyclically dominant spatial systems, this line of research is still in infancy stage. In most cases the interaction between competitors, meaning the invasion rates of a predator-prey relation, is highly symmetric, and this assumption is far from realistic models. The uniform values of the invasion rates makes analytical approaches feasible and allows a systematic scan of parameter space in simulations. But we cannot stop at this stage and should move towards more complex, hence more realistic models where heterogeneous rates are assumed. In the latter case the topology of food-web relation between competitors can not provide the full information to predict the possible solutions and may be unable to determine the vitality of emerging alliances formed by smaller group of individuals.

Because of the complexity of driving dynamics most of published works on spatial systems assumed a fixed interaction graph between competitors. There are several cases, however, when evolving graph, hence co-evolutionary modeling and off-lattice approach, are more realistic \cite{choi_epjb17,avelino_epl18}. The connection between the Hamming distance density \cite{hamming_bstj50,bazeia_epl17}, the density of maxima in open quantum dots \cite{ramos_prl11} and the chaotic evolution of the abundance of species in cyclic evolution \cite{bazeia_srep17} may also suggest new investigations on the subject. And yet, a recent work \cite{bazeia_epl20} on clustering of living species in an off-lattice model may also be further explored, concerning the possibility to describe specific properties of clusters in biological systems such as birds and fishes, among others.  We expect the present review will foster further significant steps along these paths.

\acknowledgments
This research was supported by the Hungarian National Research Fund (Grant K-120785), CAPES - Finance Code 001, Funda\c c\~ao Arauc\'aria, INCT-FCx (CNPq/FAPESP), Conselho Nacional de Desenvolvimento Cient\'\i fico e Tecnol\'ogico (CNPq, Grants 404913/2018-0; 303469/2019-6) and Para\'\i ba State Research Foundation (FAPESQ-PB, Grant 0015/2019).

\end{document}